\documentstyle[12pt]{article}

\def\[{\left\lbrack}
\def\]{\right\rbrack}

\def\({\left(}
\def\){\right)}
\def\ih{\'\i}

\title{The non-abelian BFFT formalism for the collective coordinates
quantization of the SU(2) Skyrme model}

\author{Wilson Oliveira\thanks{e-mail:wilson@fisica.ufjf.br}
 and Jorge Ananias Neto\thanks{e-mail:jorge@fisica.ufjf.br}\\
 Departamento de F\ih sica, ICE \\ Universidade Federal de Juiz de
Fora, 36036-330 \\ Juiz de Fora, MG, Brazil }

\date{}

\begin{document}

\maketitle

\begin{abstract}
\noindent 
The collective coordinates expansion of the 
Skyrme soliton particle model gives rise to the second class 
constraints. We use the non-abelian BFFT formalism to convert this
system into the one with only first class constraints. 
Choosing two different structure functions of the non-abelian 
algebra, we obtain simplified algebraic expressions for the first class 
non-abelian Hamiltonians. 
This result shows that the non-abelian BFFT 
method is, in many aspects, richer than  the abelian BFFT formalism. 
For both of the first class Hamiltonians, we derive the Lagrangians 
which lead to the new theory. When one puts the extended phase space 
variables equal to zero, the original Skyrmion Lagrangian is reproduced. 
The method of the Dirac first class constraints
is employed to quantize these two systems. We achieve the same 
spectrum, a result which confirms the consistency 
of the non-abelian BFFT formalism.
\end{abstract}

\maketitle
\newpage

\setlength{\baselineskip} {20 pt}

\section{Introduction}

\noindent Dirac quantization of the first class constrained systems
\cite{Dirac} has many attractive features. The quantum theory can be 
constructed through defining the physical states which are annihilated 
by the operators of the first class
constraints,  and then, taking the mean value of the canonical
operators, we obtain the physical values. In the resulting quantum mechanics 
there is no Dirac bracket, and consequently, one can avoid such  
difficult problems as the complicated general solution of the 
Dirac brackets and also the factor-ordering problems. This problem,
on the other hand, appears in the explicit representation of the canonical
operators. 

It is well known that the first class constraints satisfy the 
algebra \cite{HTEI}

\begin{eqnarray}
\{ T_a,T_b \} = C_{ab}^c T_c,\\
\{T_a,H_0 \}=B_a^b T_b,
\end{eqnarray}

\noindent where 
$T_a$ and $T_b$ are the first class constraints, $C_{ab}^c$
and $B_a^b$ are the structure constants and
$H_0$ is the original Hamiltonian. The physical states
are obtained by imposing the condition

\begin{equation}
\tilde{T}_\alpha | \psi \rangle_{phys} = 0, \,\,\,\, \alpha=1,2,
\end{equation}

\noindent where $\tilde{T}_\alpha$ are the operators of the first 
class constraints.

If the system has only the second class constraints then it is possible 
to convert these constraints into the ones of the first class by 
extending the phase space under special rules.
After this, one applies the Dirac procedure described above. 
Batalin, Fradkin, 
Fradkina and Tyutin\cite{BFFT} developed an elegant formalism
of transforming systems with the second class constraints into
the systems which contain only the first class constraints.
This is achieved with the aid of auxiliary fields which serve to
extend the phase space in a convenient way to transform the 
second class into first class constraints.  
This procedure is known as the BFFT formalism.
The original theory is matched when the so called unitary gauge is 
chosen.

In the original formulation of the BFFT formalism, the 
resulting first class constraints form an Abelian
algebra. This is naturally the case for the systems with the linear 
second class constraints. Recently, Banerjee, Banerjee and 
Ghosh \cite{Banerjee1}, have studied the non-abelian Proca model, and 
Oliveira and Barcelos \cite{BW} have studied the non-linear sigma model. 
In these works the BFFT formalism has been adapted in order
that the first class constraints can form a non-abelian algebra.
\footnote{For the systems with initial first and second class constraints,
the former had also to be modified in order to keep the same initial algebra, 
either abelian or non-abelian \cite{Kim}.}From these examples,
it might appear that the original formulation of the BFFT formalism is only
addressed to the theories with linear second class constraints, while the 
extension of Banerjee, Banerjee and Ghosh is addressed to the non-linear ones.
At same time the non-linear second class constraints for the same non-abelian Proca model and for the Skyrme model \cite{Skyrme} have been 
recently studied in
the context of the original BFFT formalism \cite{Banerjee3,WOJAN}. In spite
of this, it is important to emphasize that the possibility pointed out by
Banerjee, Banerjee and Ghosh that one can obtain a non-abelian first 
class theory leads to a richer structure compared with the usual BFFT case. 

The purpose of this article is to convert the second class constraints,
which arise after the collective coordinates expansion of the Skyrme model 
into first class ones. We achieve this by applying the non-abelian BFFT
formalism, and thus, employ
the Dirac method of first class constraints to quantize this system.  

The paper is organized as follows. In Sec. 2, we give a brief outline
of the usual BFFT formalism and its non-abelian extension. We also
emphasize and clarify some of the particular aspects of the formalism.
In Sec. 3, we apply the non-abelian
BFFT formalism for the collective coordinates quantization of
the SU(2) Skyrme model. We make a special choice for the
structure functions, and consequently, obtain two different 
simplified algebras for the first class constraints and the
non-abelian extended Hamiltonians. By using the Faddeev-Senjanovich 
path integral procedure \cite{FS} we derive the
Lagrangians that lead the new theories. In Sec. 4, the spectrum 
of the two simplified extended theories is calculated. In Sec. 5,
we present the conclusions.

\section {Brief review of the BFFT formalism and its non-abelian extension}

\renewcommand{\theequation}{2.\arabic{equation}}
\setcounter{equation}{0}

Let us consider a system described by a Hamiltonian $H_0$ in a
phase space $(q^i,p^i)$ with $i=1,\dots,N$. Here we suppose that the
coordinates are bosonic (extensions to include fermionic degrees of
freedom and to the continuous case can be done in a straightforward
way). It is also supposed that there the system possesses only the second class constraints. Denoting them by $T_a$, with $a=1,\dots ,M<2N$,
we arrive at the following algebra

\begin{equation}
\bigl\{T_a,\,T_b\bigr\}=\Delta_{ab},
\label{2.1}
\end{equation}

\noindent
where $\det(\Delta_{ab})\not=0$.

As it was mentioned above, the general purpose of the BFFT formalism
is to convert the second class constraints into the first class ones. 
This goal is achieved by
introducing canonical variables, one for each of the second class constraint
(the connection between the number of the second class constraints and
the number of the new variables should be equal in order to preserve
the same number of the physical degrees of freedom in the resulting extended
theory). We denote these auxiliary variables by $\eta^a$ and assume
that they satisfy the following algebra

\begin{equation}
\bigl\{\eta^a,\,\eta^b\bigr\}=\omega^{ab}.
\label{2.2}
\end{equation}

\noindent
Here $\omega^{ab}$ is a constant non-degenerate matrix
(~det$(\omega^{ab})\neq 0$~).
The obtainment of $\omega^{ab}$ is embodied in the calculation 
of the resulting first class constraints which are denoted as 
$\tilde T_a$. Of course, these constraints depend on the new
variables $\eta^a$, that is

\begin{equation}
\tilde T_a=\tilde T_a(q,p;\eta),
\label{2.3}
\end{equation}

\noindent 
and are supposed to satisfy the boundary condition

\begin{equation}
\tilde T_a(q,p;0)= T_a(q,p).
\label{2.4}
\end{equation}

\noindent
In the framework of the BFFT formalism, the characteristic property of
the new constraints is that they are assumed to be strongly
involutive, i.e.

\begin{equation}
\bigl\{\tilde T_a,\,\tilde T_b\bigr\}=0.
\label{2.5}
\end{equation}

\noindent
The solution of Eq.~(\ref{2.5}) can be achieved by considering
$\tilde T_a$ expanded as

\begin{equation}
\tilde T_a=\sum_{n=0}^\infty T_a^{(n)},
\label{2.6}
\end{equation}

\noindent
where $T_a^{(n)}$ is a term of order $n$ in $\eta$. The condition of compatibility with the boundary condition~(\ref{2.4}) requires 

\begin{equation}
T_a^{(0)}=T_a.
\label{2.7}
\end{equation}

\noindent
Substituting the Eq.(\ref{2.6}) into~(\ref{2.5}) leads to a set of
equations, one for each coefficient of $\eta^n$. We list some of them
below

\begin{eqnarray}
&&\bigl\{T_a,T_b\bigr\}
+\bigl\{T_a^{(1)},T_b^{(1)}\bigr\}_{(\eta)}=0
\label{2.8}\\
&&\bigl\{T_a,T_b^{(1)}\bigr\}+\bigl\{T_a^{(1)},T_b\bigr\}
+\bigl\{T_a^{(1)},T_b^{(2)}\bigr\}_{(\eta)}
+\bigl\{T_a^{(2)},T_b^{(1)}\bigr\}_{(\eta)}=0
\label{2.9}\\
&&\bigl\{T_a,T_b^{(2)}\bigr\}
+\bigl\{T_a^{(1)},T_b^{(1)}\bigr\}_{(q,p)}
+\bigl\{T_a^{(2)},T_b\bigr\}
+\bigl\{T_a^{(1)},T_b^{(3)}\bigr\}_{(\eta)}
\nonumber\\
&&\phantom{\bigl\{T_a^{(0)},T_b^{(2)}\bigr\}_{(q,p)}}
+\bigl\{T_a^{(2)},T_b^{(2)}\bigr\}_{(\eta)}
+\bigl\{T_a^{(3)},T_b^{(1)}\bigr\}_{(\eta)}=0
\label{2.10}\\
&&\phantom{\bigl\{T_a^{(0)},T_b^{(2)}\bigr\}_{(q,p)}+}
\vdots
\nonumber
\end{eqnarray}

\noindent 
Here the notations $\{,\}_{(q,p)}$ and $\{,\}_{(\eta)}$, represent the
parts of the Poisson bracket $\{,\}$ corresponding to the variables
$(q,p)$ and $(\eta)$, respectively. The
equations above are used iteratively to obtain the
corrections $T^{(n)}$ ($n\geq1$). Equation~(\ref{2.8}) gives
$T^{(1)}$. Using this result together with the Eq.~(\ref{2.9}), 
one calculates $T^{(2)}$, and so on. Since $T^{(1)}$ is linear 
in $\eta$ one can write it as

\begin{equation}
T_a^{(1)}=X_{ab}(q,p)\,\eta^b,
\label{2.11}
\end{equation}

\noindent where $X_{ab}$ are some new quantities.
Substituting this expression into (\ref{2.8}) and using
(\ref{2.1}) and (\ref{2.2}), we obtain

\begin{equation}
\Delta_{ab}+X_{ac}\,\omega^{cd}\,X_{bd}=0.
\label{2.12}
\end{equation}

\noindent 
We notice that this equation does not define $X_{ab}$ in a unique way,
because it also contains the still unknown elements $\omega^{ab}$. 
What is usually done is to choose $\omega^{ab}$ in such a way that the new
variables are unconstrained. One mights mention that
sometimes it is not possible to make such a choice\cite{Barc2}.
In this case, the new variables remain constrained. Consequently, the
consistency of the method requires an introduction of other new
variables in order to transform these constraints into the
first class ones. This may lead to an endless process. It is
important to emphasize that $\omega^{ab}$ can be fixed anyway.
However, even if one fixes $\omega^{ab}$, it is still not possible to
obtain a unique solution for $X_{ab}$. Let us check this point.
Since we are only considering  bosonic coordinates~\footnote{The
problem also exists for the fermionic sector.}, 
$\Delta_{ab}$ and $\omega^{ab}$ are antisymmetric quantities. So,
expression (\ref{2.12}) includes $M(M-1)/2$ independent
equations.  On the other hand, since there is no additional symmetry 
involving $X_{ab}$,  they should represent a set of $M^2$ 
independent quantities.

In the case when $X_{ab}$ does not depend on ($q,p$), it is easily
seen that the expression $T_a+\tilde T_a^{(1)}$ is already strongly 
involutive for any choice we make and we succeed in obtaining 
$\tilde T_a$. If this
is not so, the usual procedure is to introduce $T_a^{(1)}$ into Eq.
(\ref{2.9}) in order to calculate $T_a^{(2)}$ and so on. At this 
point one faces a problem that has been the origin of some developments 
of the BFFT method, including the adoption of a non-abelian constraint algebra. This occurs because we do not know {\it a priori} what is the best
choice we can make to go from one step to another. Sometimes it is
possible to figure out a convenient choice for $X_{ab}$ in order to
obtain a first class (abelian) constraint algebra at the first stage
of the process \cite{Banerjee3}. It is opportune to mention
that in ref. \cite{Banerjee4}, the use of a
non-abelian algebra was in fact a way of avoiding to dealing with the higher
orders of the iterative method. More recently, the method has been
used (in its abelian version) beyond the first correction
\cite{Banerjee2} but we mention that sometimes there are problems in
doing this \cite{Barc1}.

\newpage
Another point of the usual BFFT formalism is that any dynamic function
$A(q,p)$ (for instance, the Hamiltonian) has also to be properly
modified in order to be strongly involutive with the first class
constraints $\tilde T_a$. Denoting the modified quantity by $\tilde
A(q,p;\eta)$, we then have

\begin{equation}
\bigl\{\tilde T_a,\,\tilde A\bigr\}=0.
\label{2.13}
\end{equation}

\noindent
In addition, $\tilde A$ has to satisfy  the boundary condition

\begin{equation}
\tilde A(q,p;0)=A(q,p).
\label{2.14}
\end{equation}

\noindent The derivation of $\tilde A$ is similar to what has been 
done in getting $\tilde T_a$. Therefore, we consider an expansion 
of the form

\begin{equation}
\tilde A=\sum_{n=0}^\infty A^{(n)},
\label{2.15}
\end{equation}

\noindent 
where $A^{(n)}$ is also a term of order $n$ in $\eta$'s.
Consequently, the compatibility with Eq.~(\ref{2.14}) requires that

\begin{equation}
A^{(0)}=A.
\label{2.16}
\end{equation}

\noindent 
The combination of Eqs.~(\ref{2.6}), (\ref{2.7}), (\ref{2.13}),
(\ref{2.15}), and (\ref{2.16}) gives the equations

\begin{eqnarray}
&&\bigl\{T_a,A\bigr\}
+\bigl\{T_a^{(1)},A^{(1)}\bigr\}_{(\eta)}=0
\label{2.17}\\
&&\bigl\{T_a,A^{(1)}\bigr\}+\bigl\{T_a^{(1)},A\bigr\}
+\bigl\{T_a^{(1)},A^{(2)}\bigr\}_{(\eta)}
+\bigl\{T_a^{(2)},A^{(1)}\bigr\}_{(\eta)}=0
\label{2.18}\\
&&\bigl\{T_a,A^{(2)}\bigr\}
+\bigl\{T_a^{(1)},A^{(1)}\bigr\}_{(q,p)}
+\bigl\{T_a^{(2)},\bigr\}
+\bigl\{T_a^{(1)},A^{(3)}\bigr\}_{(\eta)}
\nonumber\\
&&\phantom{\bigl\{T_a^{(0)},A^{(2)}\bigr\}_{(q,p)}}
+\bigl\{T_a^{(2)},A^{(2)}\bigr\}_{(\eta)}
+\bigl\{T_a^{(3)},A^{(1)}\bigr\}_{(\eta)}=0
\label{2.19}\\
&&\phantom{\bigl\{T_a^{(0)},A^{(2)}\bigr\}_{(q,p)}+}
\vdots
\nonumber
\end{eqnarray}

\noindent
which correspond to the coefficients of the powers 0, 1, 2, etc$\dots$  of
the variable $\eta$. It is just a matter of algebraic
work to show that the general expression for $A^{(n)}$ reads as

\begin{equation}
A^{(n+1)}=-{1\over n+1}\,\eta^a\,\omega_{ab}\,X^{bc}\,G_c^{(n)}.
\label{2.20}
\end{equation}

\noindent 
where $\omega_{ab}$ and $X^{ab}$ are the inverses of $\omega^{ab}$
and $X_{ab}$, and

\begin{eqnarray}
G_a^{(n)}=\sum_{m=0}^n\bigl\{T_a^{(n-m)},\,A^{(m)}\bigr\}_{(q,p)}
+\sum_{m=0}^{n-2}\bigl\{T_a^{(n-m)},\,A^{(m+2)}\bigr\}_{(\eta)}\nonumber\\
+\bigl\{T_a^{(n+1)},\,A^{(1)}\bigr\}_{(\eta)}.
\label{2.21}
\end{eqnarray}

\noindent The general prescription of the usual BFFT method to obtain the
Hamiltonian is a direct use of the relations (\ref{2.15}) and
(\ref{2.20}). This works well for the system with linear constraints. For
non-linear theories, where it may be necessary to consider all orders
of the iterative process, this calculation might be quite
complicated. However, there is an alternative procedure that drastically
simplifies the algebraic work. The basic idea is to
obtain the involutive forms for the initial fields $q$ and $p$
\cite{Banerjee5}. This can be directly achieved from the previous
calculation of $\tilde A$. Denoting such fields by $\tilde q$ and
$\tilde p$ we have

\begin{equation}
H(q,p)\longrightarrow H(\tilde q,\tilde p)
=\tilde H(\tilde q,\tilde p).
\label{2.22}
\end{equation}

\noindent
It is obvious that the initial boundary condition in the BFFT
process, that is, the reduction of the involutive function to the
original function when the new fields are set to zero, remains
preserved. One can also mention that for the systems with linear
constraints, the new variables $\tilde q$ and $\tilde p$ are just
shifted in the auxiliary coordinate $\eta$ \cite{Ricardo}.

Finally, let us consider the case where the first class
constraints form a non-abelian algebra, i.e. 

\begin{equation}
\bigl\{\tilde T_a,\,\tilde T_b\bigr\}=C_{ab}^c\,\tilde T_c.
\label{2.23}
\end{equation}

\noindent
The quantities $C_{ab}^c$ are the structure constants of the
non-abelian algebra. These constraints are considered to satisfy the
same previous conditions given by (\ref{2.3}), (\ref{2.4}),
(\ref{2.6}), and (\ref{2.7}). But now, instead of Eqs.
(\ref{2.8})-(\ref{2.10}), we obtain 

\begin{eqnarray}
C_{ab}^c\,T_c&=&\bigl\{T_a,T_b\bigr\}
+\bigl\{T_a^{(1)},T_b^{(1)}\bigr\}_{(\eta)}
\label{2.24}\\
C_{ab}^c\,T_c^{(1)}&=&\bigl\{T_a,T_b^{(1)}\bigr\}
+\bigl\{T_a^{(1)},T_b\bigr\}
\nonumber\\
&&+\,\bigl\{T_a^{(1)},T_b^{(2)}\bigr\}_{(\eta)}
+\bigl\{T_a^{(2)},T_b^{(1)}\bigr\}_{(\eta)}
\label{2.25}\\
C_{ab}^c\,T_c^{(2)}&=&\bigl\{T_a,T_b^{(2)}\bigr\}
+\bigl\{T_a^{(1)},T_b^{(1)}\bigr\}_{(q,p)}
\nonumber\\
&&+\bigl\{T_a^{(2)},T_b^{(0)}\bigr\}_{(q,p)}
+\bigl\{T_a^{(1)},T_b^{(3)}\bigr\}_{(\eta)}
\nonumber\\
&&+\bigl\{T_a^{(2)},T_b^{(2)}\bigr\}_{(\eta)}
+\bigl\{T_a^{(3)},T_b^{(1)}\bigr\}_{(\eta)+}
\label{2.26}\\
&&\vdots
\nonumber
\end{eqnarray}

\noindent 
The use of these equations is the same as before, i.e., they shall
work iteratively. Equation (\ref{2.24}) gives $T^{(1)}$.  With this
result and Eq. (\ref{2.25}) one calculates $T^{(2)}$, and so on. To
calculate the first correction, we assume it is given by the same
general expression (\ref{2.11}). Introducing it into (\ref{2.24}), we
now get

\begin{equation}
C_{ab}^c\,T_c=\Delta_{ab}+X_{ac}\,\omega^{cd}\,X_{bd}.
\label{2.27}
\end{equation}

\noindent 
Of course, the same difficulties concerning the
solutions of Eq.~(\ref{2.12}) also apply here, with the additional
problem of choosing the appropriate structure constants $C_{ab}^c$.
To obtain the embedding Hamiltonian $\tilde H(q,p,\eta)$ one cannot
use the simplified version discussed for the abelian case (embodied
into Eq.~(\ref{2.22}) ) because the algebra is not strong involutive
anymore. Thus we start from the fact that the new Hamiltonian $\tilde
H$ and the new constraints $\tilde T_a$ satisfy the relation

\begin{equation}
\bigl\{\tilde T_a,\,\tilde H\bigr\}=B_a^b\,\tilde T_b,
\label{2.28}
\end{equation}

\noindent where the coefficients $B_a^b$ are the 
structure constant of the non-abelian algebra. The involutive 
Hamiltonian is considered to 
satisfy the same conditions\break(\ref{2.14})-(\ref{2.16}). We then obtain
that the general correction $H^{(n)}$ is given by a relation similar
to (\ref{2.20}), but now the quantities $G_a^{(n)}$ are given by

\begin{eqnarray}
G_a^{(n)}&=&\sum_{m=0}^n\bigl\{T_a^{(n-m)},\,H^{(m)}\bigr\}_{(q,p)}
+\sum_{m=0}^{n-2}\bigl\{T_a^{(n-m)},\,A^{(m+2)}\bigr\}_{(\eta)}
\nonumber\\
&&+\,\,\bigl\{T_a^{(n+1)},\,A^{(1)}\bigr\}_{(\eta)}
-B_a^b\,T_c^{(n)}.
\label{2.30}
\end{eqnarray}

\section {The non-abelian BFFT formalism for the SU(2) Skyrme model}

 The classical static Lagrangian of the Skyrme model
is given by

\begin{equation}
\label{clag}
L = \int d^3r \{ -{F_\pi^2\over 16} Tr \(\partial_i U
\partial_i U^+ \) + {1\over 32 e^2} Tr \[ U^+\partial_i U,
U^+ \partial_j U \]^2 \} \, ,
\end{equation}

\noindent where $F_\pi$ is the pion decay constant, {\it e}
is a dimensionless parameter and U is an SU(2) matrix.
Performing the collective semi-classical expansion\cite{ANW},
substituting U(r) by $U(r,t)=A(t)U(r)A^+ (t)$ in (\ref{clag}),
where A is an SU(2) matrix, we obtain

\begin{equation}
\label{Lag}
L = - M + \lambda Tr [ \partial_0 A\partial_0 A^{-1} ],
\end{equation}

\noindent where M is the soliton mass. In the hedgehog
representation for U, $U=\exp(i\tau \cdot \hat{r} F(r))$, 
this mass is given by

\begin{equation}
\label{henergia}
M = 4\pi {F_\pi \over e} \int^\infty_0 dx { x^2 {1\over 8}
\[ F'^2 + 2 {2 \sin^2 F \over x^2} \]  + {1\over 2}
\[ {\sin^2 F\over x^2} + 2 F'^2 \] } ,
\end{equation}

\noindent where {\it x} is a dimensionless variable defined
by $x=eF_\pi r$ and $\lambda$ is called the inertia moment
written as

\begin{equation}
\label{lambda}
\lambda = {2\over 3} \pi ({1\over e^3 F_\pi}) \Lambda
\end{equation}

\noindent with

\begin{equation}
\label{Lambda}
\Lambda = \int^\infty_0 dx x^2 \sin^2F \[ 1 +
4(F'^2 + {\sin^2 F\over x^2}) \].
\end{equation}

\noindent The SU(2) matrix A can be written as $A=a^0
+i a\cdot \tau$ with the constraint

\begin{equation}
\label{pri}
T_1 = a^ia^i - 1 \approx 0, \,\,\,\, i=0,1,2,3.
\end{equation}

\noindent The Lagrangian(\ref{Lag}) can be written as a function of the
$a^i$ as

\begin{equation}
\label{cca}
L = -M + 2\lambda \dot{a}^i\dot{a}^i.
\end{equation}

\noindent In order to identify more constraints, we calculate the
momentum

\begin{equation}
\label{cm}
\pi^i = {\partial L \over \partial \dot{a}_i} = 4 \lambda \dot{a}^i.
\end{equation}

\noindent Now we can  rewrite the Hamiltonian in the form

\begin{eqnarray}
\label{chr}
H_c=\pi^i \dot a^i-L=4\lambda \dot a^i \dot a^i -L=M+2
 \lambda \dot a^i\dot a^i\nonumber\\
=M+{1\over 8 \lambda } \sum_{i=0}^3 \pi^i\pi^i.
\end{eqnarray}

\noindent Constructing the total Hamiltonian and imposing the
consistency condition that constraints do not evolve in time
\cite{Dirac} we get a new constraint

\begin{equation}
\label{T2}
T_2 = a^i\pi^i \approx 0 \,\,.
\end{equation}

\noindent We observe that no further constraints are generated
via this iterative procedure. The constraints $T_1$ and $T_2$
are of the second class. The matrix elements of their Poisson 
brackets read

\begin{equation}
\label{Pa}
\Delta_{\alpha \beta} = \{T_\alpha,T_\beta\} = -2 \epsilon_{\alpha \beta}
a^ia^i, \,\, \alpha,\beta = 1,2
\end{equation}

\noindent where $\epsilon_{\alpha \beta}$ is the antisymmetric
tensor normalized as $\epsilon_{12} = -\epsilon^{12} = -1$.
\par Then, the standard quantization is made where we replace
$\,\pi^i \,$ by $\, -i \partial/\partial a_i \,$ in (\ref{chr}),
leading to

\begin{equation}
\label{uqh}
H=M+{1\over 8 \lambda } \sum_{i=0}^3 (-{\partial^2
\over\partial{a_i}^2})\,\,.
\end{equation}

\noindent Due the constraint $\sum_{i=0}^3a^i a^i=1$, 
the operator $\sum_{i=0}^3 (-{\partial^2\over\partial{a_i}^2})$
must be interpreted as the Laplacian on the three-sphere\cite{ANW}.
A typical polynomial wave function\cite{ANW},
${1\over N(l)}(a^1 + i a^2)^l = |polynomial \rangle\, ,$ is an 
eigenvector of the Hamiltonian
(\ref{uqh}), with the eigenvalues given by \footnote
{This wave function is also eigenvector of the spin and
isospin operators, written as\cite{ANW} $ J^k={1\over 2}
( a_0 \pi_k -a_k \pi_0 - \epsilon_{klm} a_l \pi_m )$  and 
$ I^k={1\over 2 } ( a_k \pi_0 -a_0 \pi_k- \epsilon_{klm} a_l
\pi_m ).$}.

\begin{equation}
\label{uqhe}
E=M+{1\over 8 \lambda } l(l+2), \,\,\,\, l=1,2,3\dots \,\,.
\end{equation}

\vskip .5cm

 To implement the extended non-abelian BFFT formalism, we introduce 
 auxiliary coordinates, one for each of the second class constraint.
Let us generally denote them by $\eta^\alpha$, where $\alpha=1,2$, 
and consider that the Poisson algebra of these new coordinates 
is given by

\begin{equation}
\label{algebra1}
\{ \eta^\alpha, \eta^\beta \} = \omega^{\alpha \beta}
= 2\epsilon^{\alpha\beta}; 
\,\,\alpha=1,2.
\end{equation}

\noindent From Eq.~(\ref{2.27}), we have

\begin{equation}
2X_{11}(x,z)\,X_{22}(y,z) = -2\,a^i a^i\ + \,C_{12}^1\,T_1.
\label{3.15}
\end{equation}

\noindent
After some attempts, we find that a convenient choice for these 
coefficients is

\begin{eqnarray}
&&X_{11}=1,
\nonumber\\
&&X_{22}=-1,
\nonumber\\
&&X_{12}=0=X_{21},
\nonumber\\
&&C_{12}^1=2,
\nonumber\\
&&C_{12}^2=0.
\label{3.16}
\end{eqnarray}

\noindent Using (\ref{2.4}), (\ref{2.6}), (\ref{2.11}), (\ref{algebra1}) and
(\ref{3.16}), the new set of constraints is found to be

\begin{eqnarray}
\label{TF1}
\tilde{T}_1=a^i a^i-1+\eta^1,\\
\label{TF2}
\tilde{T}_2=a^i\pi^i-\eta^2+\eta^1\eta^2.
\end{eqnarray}

\noindent The first class constraint algebra is

\begin{eqnarray}
&&\bigl\{\tilde T_1,\,\tilde T_1\bigr\}=0,
\nonumber\\
&&\bigl\{\tilde T_1,\,\tilde T_2\bigr\}
=2\,\tilde T_1,
\nonumber\\
&&\bigl\{\tilde T_2,\,\tilde T_2\bigr\}=0.
\label{3.24}
\end{eqnarray}

Next, we derive the corresponding Hamiltonian in the extended
phase space. The corrections for the canonical Hamiltonian are
given by Eqs. (\ref{2.20}) and (\ref{2.30}). With the objective
to simplify the expression of the first class Hamiltonian, we
chose two different algebras for the system defined by the parameters
$B_a^b$ in (\ref{2.28}). We have verified that possible values are

\begin{equation}
\label{sys1}
B_a^b=0, \,\,\,\, a,b=1,2,
\end{equation}

\noindent and

\begin{equation}
\label{sys2}
B_1^1={1\over 2\lambda},\,\,\,\, B_1^2=B_2^1=B_2^2=0.
\end{equation}

\noindent Using the inverse matrices

\begin{eqnarray}
\label{apc1}
\omega_{\alpha\beta} = {1\over 2} \epsilon_{\alpha\beta}, \\\
\label{apc2}
X^{\alpha \beta} = \left( \begin{array}{clcr} 1  & \,\,0 \\ 0
& -1\end{array} \right),
\end{eqnarray}

\noindent and the algebra defined by (\ref{sys1}), it is possible
to compute the involutive first class Hamiltonian

\newpage

\begin{eqnarray}
\label{HF1F}
\tilde{H}=M + {1\over 8\lambda} \pi^i\pi^i
-{1\over 8\lambda} \pi^i\pi^i \eta^1
-{1\over 4\lambda} a^i\pi^i\eta^2
+{1\over 4\lambda} a^i\pi^i \eta^1\eta^2\nonumber\\
+ {1\over 8\lambda} a^i a^i\eta^2\eta^2
-{1\over 8\lambda} a^i a^i\eta^1\eta^2\eta^2\nonumber\\\nonumber\\
=M+{1\over 8\lambda} \pi^i\pi^i(1-\eta^1)
-{1\over 4\lambda} a^i\pi^i\eta^2(1-\eta^1)
+ {1\over 8\lambda} a^i a^i\eta^2\eta^2(1-\eta^1).
\end{eqnarray}

\noindent Thus, the Hamiltonian (\ref{HF1F}) satisfies the first class algebra

\begin{eqnarray}
\label{HF11}
\{ \tilde{T}_1, \tilde{H}_1 \} = 0,\,\,\,\,\,\,\,(B_1^1=B_1^2=0)\\
\label{HF21}
\{ \tilde{T}_2, \tilde{H}_1 \} = 0. \,\,\,\,\,\,\,(B_2^1=B_2^2=0)
\end{eqnarray}

\noindent The other non-abelian  first class Hamiltonian is given by

\begin{eqnarray}
\label{HF2F}
\tilde{H}_2=\tilde{H}_1+{1\over 4\lambda} \tilde{T}_2\nonumber\\\nonumber\\
=M+{1\over 8\lambda} \pi^i\pi^i(1-\eta^1)
-{1\over 4\lambda} a^i\pi^i\eta^2(1-\eta^1)
+ {1\over 8\lambda} a^i a^i\eta^2\eta^2(1-\eta^1)\nonumber\\
+{1\over 4\lambda} (a^i\pi^i-\eta^2(1-\eta^1)),
\end{eqnarray}

\noindent which satisfies the first class Poisson algebra

\begin{eqnarray}
\label{HF12}
&\{\tilde{T}_1, \tilde{H}_2\} = {1\over 2\lambda}\tilde{T}_1,\,\,
\,\,\,\,(B_1^1={1\over 2\lambda},B_1^2=0)\\
\label{HF22}
&\{\tilde{T}_2, \tilde{H}_2 \} = 0. \,\,\,\,\,\,\,\,\,\,\,\,\,\,
\,\,\,\,\,\,\,\,
(B_2^1=B_2^2=0)
\end{eqnarray}

\newpage

\noindent Here we would like to remark that, contrary the results
obtained by the abelian BFFT method applied to the non-linear
Lagrangian theories \cite{Banerjee3,WOJAN}, both expressions of the
first class Hamiltonians (\ref{HF1F}) and (\ref{HF2F}) are finite
sums. As it was emphasized in the introduction, the possibility 
pointed out by Banerjee, Banerjee and Ghosh to obtain non-abelian
first class theories leads to a  more elegant and  simplified 
Hamiltonian structure than usual abelian BFFT case.

   The next step is to look for the Lagrangian that leads
to this new theory. A consistent way of doing this is by
means of the  path integral formalism, where the Faddeev
procedure \cite{FS} has to be used. Let us identify the new 
variables $\eta^\alpha$ as a canonically conjugate pair $ (\phi, 
\pi_\phi)$ in the Hamiltonian formalism,

\begin{eqnarray}
\label{cpair}
\eta^1 \rightarrow 2 \phi \,, \nonumber \\
\eta^2 \rightarrow \pi_\phi \,,
\end{eqnarray}

\noindent satisfying (\ref{algebra1}). 
Then, the general expression for the vacuum functional reads

\begin{equation}
\label{vfg}
Z = N \int [d\mu] \exp \{ i \int dt [ \dot{a}^i\pi^i
+ \dot{\phi}\pi_\phi - \tilde{H} ] \},
\end{equation}

\noindent with the measure $[d\mu]$ given by

\begin{eqnarray}
\label{mesure}
[d\mu] = [da^i] [d\pi^i] [d\phi] [d\pi_\phi]
| det\{,\} | \nonumber \\ \delta(a^ia^i-1+2\phi)
\delta(a^i\pi^i- \pi_\phi +2\phi\pi_\phi)\prod_\alpha
\delta(\tilde{\Lambda}_\alpha),
\end{eqnarray}

\noindent where $\tilde{\Lambda}_\alpha$ are the gauge fixing 
conditions corresponding to the first class constraints 
$\tilde{T}_\alpha$ and the term $| det\{,\} |$ represents
the determinant of all constraints of the theory, including
the gauge-fixing ones. The quantity N that appears in 
(\ref{vfg}) is an usual normalization factor.  Starting from
the Hamiltonian (\ref{HF1F}), the vacuum functional reads

\begin{eqnarray}
\label{pf1}
Z = N \int [da^i] [d\pi^i] [d\phi] [d\pi_\phi] 
| det\{,\} | \, \delta( a^ia^i-1+2\phi )\nonumber\\
\delta(a^i\pi^i-\pi_\phi(1-2\phi))
\prod_\alpha
\delta(\tilde{\Lambda}_\alpha)\exp \{ i \int dt 
[ \dot{a}^i\pi^i + \dot{\phi}\pi_\phi
-M\nonumber\\
-{1\over 8\lambda} \pi^i\pi^i(1-2\phi)
+{1\over 4\lambda} a^i\pi^i \pi_\phi(1-2\phi)
-{1\over 8\lambda} a^i a^i \pi_\phi\pi_\phi(1-2\phi)]\}.
\end{eqnarray}

\noindent Using the delta function $\delta(a^ia^i-1+2\phi)$ and
exponentiating the delta function $\delta[a^i\pi^i-\pi_\phi(1-2\phi)]$
with Fourier variable $\xi$, we obtain

\begin{eqnarray}
\label{ele}
Z = N \int [da^i] [d\pi^i] [d\phi] [d\pi_\phi] [d\xi]
| det\{,\} | \, \delta( a^ia^i-1+2\phi ) \prod_\alpha
\delta(\tilde{\Lambda}_\alpha) \nonumber \\ \exp \{ i \int dt 
[ \dot{a}^i\pi^i + \dot{\phi}\pi_\phi
-M
-{1\over 8\lambda}\pi^j\pi^j a^ia^i
+{1\over 4\lambda} a^j\pi^j a^ia^i \pi_\phi\nonumber\\
-{1\over 8\lambda} (a^i a^i)^2 (\pi_\phi)^2
+\xi a^i\pi^i-\xi a^i a^i \pi_\phi
]\}.
\end{eqnarray}

\noindent Integrating over $\pi_\phi$, we arrive at,

\begin{eqnarray}
\label{pifi}
Z = N \int [da^i] [d\pi^i] [d\phi] [d\xi]
| det\{,\} | \, \delta( a^ia^i-1+2\phi ) \prod_\alpha
\delta(\tilde{\Lambda}_\alpha) \nonumber \\ 
{1\over a^ia^i}\exp \{ i \int dt 
[ \dot{a}^i\pi^i 
-M
-{1\over 8\lambda}\pi^j\pi^j a^i a^i
+ \xi a^i\pi^i\nonumber\\
-{2\lambda \dot{\phi}\dot{\phi}\over {(a^ia^i)}^2}
+ {4\lambda \dot{\phi}\xi\over a^ia^i}
- 2\lambda\xi^2
 ]\}.
\end{eqnarray}

\newpage

\noindent Performing the integration over $\pi^i$, we obtain

\begin{eqnarray}
\label{pii}
Z = N \int [da^i][d\phi] [d\xi]
| det\{,\} | \, \delta( a^ia^i-1+2\phi ) \prod_\alpha
\delta(\tilde{\Lambda}_\alpha) \nonumber \\ 
{1\over 1-2\phi}
\sqrt{{1\over 1-2\phi}}
\exp \{ i \int dt 
[ 
-M 
+{2\lambda}{ \dot{a}^i \dot{a}^i\over 1-2\phi}\nonumber \\
- {2\lambda}{\dot{\phi} \dot{\phi}\over (1-2\phi)^2}
+{4\lambda \over 1-2\phi}(a^i \dot{a}^i + \dot{\phi})\xi
]\}.
\end{eqnarray}

\noindent Finally, the integration over $\xi$ leads to

\begin{eqnarray}
\label{xi1}
Z = N \int [da^i][d\phi] 
|det\{,\}| \, \delta(a^ia^i-1+2\phi) \, \delta(a^i \dot{a}^i + \dot{\phi}) \prod_\alpha
\delta(\tilde{\Lambda}_\alpha) \nonumber \\ 
\sqrt{1\over 1-2\phi}
\exp  \{ i \int dt [
-M 
+{2\lambda\over 1-2\phi} \dot{a}^i\dot{a}^i\nonumber\\
-{2\lambda\over {(1-2\phi)}^2} \dot{\phi}\dot{\phi}
 ] \},
\end{eqnarray}

\noindent where the new $\delta$ function above came from the integration over $\xi$.
We notice that it is nothing other than the derivative of constraint $\tilde{T}_1$. It is then just a consistency condition and does not represent any new restriction over the coordinates of the theory. From the vacuum functional (\ref{xi1}), we identify the Lagrangian of the new theory 

\begin{equation}
\label{L1f}
L = -M 
+{2\lambda\over 1-2\phi} \dot{a}^i\dot{a}^i\nonumber\\
-{2\lambda\over {(1-2\phi)}^2} \dot{\phi}\dot{\phi}.
\end{equation}

\noindent Putting the extended variables, in the phase space,
$\phi$ and $\pi_\phi$ equal to zero, we obtain the original 
Skyrmion Lagrangian. This result indicates the consistency
of the theory.
\par  For the Hamiltonian (\ref{HF2F}) the vacuum functional is

\begin{eqnarray}
\label{ele2}
Z = N \int [da^i] [d\pi^i] [d\phi] [d\pi_\phi] 
| det\{,\} | \, \delta( a^ia^i-1+2\phi )\nonumber\\
\delta(a^i\pi^i-\pi_\phi(1-2\phi))
\prod_\alpha
\delta(\tilde{\Lambda}_\alpha)\exp \{ i \int dt 
[ \dot{a}^i\pi^i + \dot{\phi}\pi_\phi
-\tilde{H}_1-{1\over 4\lambda}\tilde{T}_2 \}
\end{eqnarray}

\noindent Using the properties of delta functions, it is easy to see
that we have obtained the same Lagrangian (\ref{L1f}).

\section{The spectrum of the theory}

\par Here we intend to obtain the spectrum of the extended theory.
We use the Dirac method of quantization for the first class constraints
\cite{Dirac}.The basic idea consists in imposing quantum 
mechanically the first class constraints as operator 
condition on the wave-functions as a way to obtain the physical 
subspace, i.e.,

\begin{equation}
\label{qope}
\tilde{T}_\alpha | \psi \rangle_{phys} = 0, \,\,\,\, \alpha=1,2.
\end{equation}

\noindent The operators $\tilde{T}_1\,$ and $\tilde{T}_2\,$
are

\begin{eqnarray}
\label{qope1}
\tilde{T}_1=a^ia^i-1+\eta^1,\\
\label{qope2}
\tilde{T}_2=a^i\pi^i - \eta^2+\eta^1\eta^2.
\end{eqnarray}

\noindent Thus, the physical states that satisfy (\ref{qope}) are

\begin{equation}
\label{physical}
| \psi \rangle_{phys} = {1\over V } \, \delta (a^i\pi^i
- \eta^2+\eta^1\eta^2) \,\delta(a^i a^i-1+\eta^1)\,|polynomial \rangle,
\end{equation}

\noindent where {\it V } is the normalization factor 
and the ket {\it polynomial} was defined in Section 3 as,
$|polynomial \rangle ={1\over N(l)} (a^1+ i a^2)^l \,$. The 
corresponding quantum Hamiltonians of (\ref{HF1F}) and (\ref{HF2F})
will be indicated as 

\begin{eqnarray}
\label{echs1}
\tilde{H_1}= M+{1\over 8\lambda} \pi^i\pi^i(1-\eta^1)
-{1\over 4\lambda} a^i\pi^i\eta^2(1-\eta^1)\nonumber\\
+ {1\over 8\lambda} a^i a^i\eta^2\eta^2(1-\eta^1),
\end{eqnarray}

\noindent and

\begin{eqnarray}
\label{echs2}
\tilde{H_2}= \tilde{H_1}+{1\over 4\lambda}\tilde{T_2}.
\end{eqnarray}

\noindent Thus, in order to obtain the spectrum of the theory, we take
 the scalar product, 
$_{phys}\langle\psi| \tilde{H} | \psi \rangle_{phys}\,$,
that is the mean value of the extended Hamiltonian, for the two
quantum Hamiltonians (\ref{echs1}) and (\ref{echs2}). We begin
with the first Hamiltonian (\ref{echs1}) calculating the scalar
product

\begin{eqnarray}
\label{mes1}
_{phys}\langle\psi| \tilde{H_1} | \psi \rangle_{phys}=\nonumber \\
\langle polynomial |\,\,  {1\over V^2} \int d\eta^1 d\eta^2 
\delta(a^i a^i - 1 + \eta^1)\delta(a^i\pi^i - \eta^2+\eta^1
\eta^2)\nonumber \\
\tilde{H}_1
\delta([a^i\pi^i - \eta^2+\eta^1\eta^2)\delta(a^i a^i - 1 + \eta^1)\,\,
| polynomial \rangle .
\end{eqnarray}

\noindent Notice that due to the presence of the delta functions
 $\delta(a^i a^i - 1 + \eta^1)$ and
$\delta(a^i\pi^i - \eta^2+\eta^1\eta^2)$ in
(\ref{mes1}) the scalar product can be simplified. 
Then, integrating over $\eta^1$ and $\eta^2$ we obtain\footnote{The 
regularization of delta function squared like
$(\delta(a^i a^i - 1 + \eta^1))^2$ and 
$(\delta(a^i\pi^i - \eta^2+\eta^1\eta^2))^2$ 
is performed by using the delta relation, $(2\pi)^2\delta(0)=
\lim_{k\rightarrow 0}\int d^2x \,e^{ik\cdot x} =\int d^2x= V.$ 
Then, we use the parameter V as the normalization factor.}

\begin{eqnarray}
\label{mes13}
_{phys}\langle\psi| \tilde{H_1} | \psi \rangle_{phys}=\nonumber \\
\langle polynomial | M + {1\over 8\lambda}  a^ia^i \pi^j \pi^j 
- {1\over 8\lambda} a^i\pi^i a^j\pi^j | polynomial \rangle .
\end{eqnarray}

\noindent We repeat the same procedure for the quantum Hamiltonian
(\ref{echs2}). Taking the mean value, we have

\begin{eqnarray}
\label{mes2}
_{phys}\langle\psi| \tilde{H_2} | \psi \rangle_{phys}=\nonumber \\
\langle polynomial |\,\,  {1\over V^2} \int d\eta^1 d\eta^2 
\delta(a^i a^i - 1 + \eta^1)\delta(a^i\pi^i - \eta^2+\eta^1
\eta^2)\nonumber \\
\tilde{H_2}
\delta([a^i\pi^i - \eta^2+\eta^1\eta^2)\delta(a^i a^i - 1 + \eta^1)\,\,
| polynomial \rangle .
\end{eqnarray}

\noindent Using the delta properties, we obtain the simplified
scalar product for the Hamiltonian (\ref{echs2})

\begin{eqnarray}
\label{mes21}
_{phys}\langle\psi| \tilde{H_2} | \psi \rangle_{phys}=\nonumber \\
\langle polynomial | M + {1\over 8\lambda}  a^ia^i \pi^j \pi^j 
- {1\over 8\lambda} a^i\pi^i a^j\pi^j | polynomial \rangle .
\end{eqnarray}

\noindent The expression above is the same obtained for the scalar 
product of quantum Hamiltonian $\tilde{H_1}$. It is important to 
remark that, despite the BFFT formalism permits to have freedom to 
choose different first class algebras for the same second class
Hamiltonian, the two expressions for the spectrum of both algebras are
identical. This result shows again the consistency of the BFFT
formalism.

   The final Hamiltonian operator inside the kets (\ref{mes13})
and (\ref{mes21}) must be hermitian. Then, this Hamiltonian has to be 
symmetrized\footnote{In the BFFT formalism  applied on the Skyrme model, the operator ordering problem appears in the expression of the non-abelian first class Hamiltonian.}. Following the prescription of Weyl ordering\cite{Weyl}
(symmetrization procedure) we can write the symmetric 
Hamiltonian as

\begin{eqnarray}
\label{HWeyl}
\tilde{H}_{sym} = {1\over 8\lambda}  \[ a^ia^i \pi^j \pi^j \]_{sym} 
- {1\over 8\lambda}\[ a^i\pi^i a^j\pi^j \]_{sym},
\end{eqnarray}

\noindent where $\[ a^ia^i \pi^j \pi^j \]_{sym} $ and
 $\[ a^i\pi^i a^j\pi^j \]_{sym}$ are defined as

\begin{eqnarray}
\label{wdef}
\[ a^i a^i \pi^j \pi^j \]_{sym} = {1\over 32\lambda}
\[ a^i( a^i \pi^j + \pi^j a^i )\pi^j + \pi^j 
( a^i \pi^j + \pi^j a^i ) a^i \]\\
\[ a^i\pi^i a^j\pi^j \]_{sym} = {1\over 32\lambda} 
\[(a^i \pi^i+ \pi^i a^i) (a^j \pi^j + \pi^j a^j) \].
\end{eqnarray}

\noindent Then, using the symmetric Hamiltonian operator 
$\tilde{H}_{sym}$ Eq.~(\ref{HWeyl}) both the mean values 
(\ref{mes13}) and (\ref{mes21}) are

\begin{eqnarray}
\label{mes2W}
_{phys}\langle\psi| \tilde{H}_{sym} | \psi \rangle_{phys}=\nonumber \\
\langle polynomial | M + {1\over 8\lambda}  \[ a^ia^i \pi^j \pi^j \]_{sym} 
- {1\over 8\lambda}\[ a^i\pi^i a^j\pi^j \]_{sym}| polynomial \rangle .
\nonumber\\
=\langle polynomial | M + {1\over 32\lambda}
[ a^i ( a^i \pi^j + \pi^j a^i )\pi^j + \pi^j 
( a^i \pi^j + \pi^j a^i ) a^i ]\nonumber \\- {1\over 32\lambda} 
[(a^i \pi^i+ \pi^ia^i) (a^j \pi^j + \pi^j a^j) ]
| polynomial \rangle \,\,.
\end{eqnarray}

\noindent The operator $\pi^j$ describes a free particle 
and its representation on the collective coordinates space $a^i$ 
is given by

\begin{equation}
\label{piconfig}
\pi^j = -i {\partial\over \partial a_j}\,\,.
\end{equation}

\noindent Substituting the expression (\ref{piconfig}) into 
(\ref{mes2W}), we obtain

\begin{equation}
\label{meswe}
_{phys}\langle\psi| [\tilde{H}]_{sym} | \psi \rangle_{phys} = 
M + {1\over 8\lambda} \[ l(l+2) + 1 \] \,\,.
\end{equation}

\noindent This last expression, Eq.~(\ref{meswe}), differs from
the conventional energy eigenvalues of the Skyrme model, Eq.~(\ref{uqhe}),
by an additional constant term. 
Thus, using the symmetrized non-abelian BFFT Hamiltonians and employing
the Dirac quantization method of first class constraints, we have
obtained the Skyrmion rotational mode energy
eigenvalues with a mass shift. Similar results have been also 
obtained by many authors\footnote{These authors have pointed out that a mass
shift can improve the usual phenomenology predicted by the Skyrme
model.}\cite{Fujii} using different procedures.

\section{Conclusions}

We have used an extension of the BFFT formalism presented by Banerjee,
Banerjee and Ghosh in order to quantize the SU(2) Skyrme model. Using
the non-abelian algebra, we have shown that, contrary to the 
results obtained by the usual abelian BFFT formalism, it is possible 
to construct the first class Hamiltonians that are simple finite sums. 
The extended Lagrangians were achieved by using the Faddeev-Senjanovich
constraint path integral formalism. In the so called unitary gauge
we reproduced the original Skyrmion Lagrangian. We calculate the 
mean energy for the two different first class 
Hamiltonian operators leading consistently to the same mass spectrum of 
the theory. Then, our results show, 
in some sense, that for the non-linear theory the non-abelian BFFT 
formalism is more adequate than the abelian formalism.

\section{Acknowledgments}
We would like to thank Ilya Shapiro for critical reading.
 This work is supported in part by FAPEMIG, Brazilian Research Council.

\end{document}